\documentclass{article}
\usepackage{geometry}
\usepackage{booktabs}
\usepackage{hyperref}
\hypersetup{colorlinks=true,linkcolor=black,citecolor=black,urlcolor=black}
\usepackage{xurl}
\usepackage{footnote}
\usepackage{longtable}
\usepackage{array}
\usepackage{multirow}
\usepackage{authblk}
\usepackage{adjustbox}

\title{The CryptoNeo Threat Modelling Framework (CNTMF): Securing Neobanks and Fintech in Integrated Blockchain Ecosystems}
\author[1]{Serhan W. Bahar}
\affil[1]{Independent Researcher, London, United Kingdom}
\date{July 2025}

\begin{document}

\maketitle

\section*{Abstract}

The rapid integration of blockchain, cryptocurrency, and Web3 technologies into digital banks and fintech operations has created an integrated environment blending traditional financial systems with decentralised elements. This paper introduces the CryptoNeo Threat Modelling Framework (CNTMF), a proposed framework designed to address the risks in these ecosystems, such as oracle manipulation and cross-chain exploits. CNTMF represents a proposed extension of established methodologies like STRIDE, OWASP Top 10, NIST frameworks, LINDDUN, and PASTA, while incorporating tailored components including Hybrid Layer Analysis, the CRYPTOQ mnemonic for cryptocurrency-specific risks, and an AI-Augmented Feedback Loop. Drawing on real-world data from 2025 incidents, CNTMF supports data-driven mitigation to reduce losses, which totalled approximately \$2.47 billion in the first half of 2025 across 344 security events (CertiK via GlobeNewswire, 2025; Infosecurity Magazine, 2025).\footnote{Note: Figures vary by source; CertiK via GlobeNewswire (2025) reports \$2.47 billion (no incident count specified), Infosecurity Magazine (2025) reports \$2.47 billion across 344 incidents, while Bitget (2025) focuses on \$2.1 billion (75 major hacks, CryptoSlate, 2025, citing TRM Labs; $\sim$80\% infrastructure, $\sim$70\% of stolen value North Korea-linked). Net losses: \$2.29 billion (CertiK).} Its phases guide asset mapping, risk profiling, prioritisation, mitigation, and iterative feedback. This supports security against evolving risks like state-sponsored attacks.

\section{Introduction}

The integration of blockchain, cryptocurrency, and Web3 technologies into digital banks and fintech companies has accelerated. It enables innovative services such as seamless fiat on/off-ramps, decentralised finance (DeFi) protocols for lending and yield farming, and multi-party computation (MPC) wallets for enhanced key management and security (Deloitte, 2025). This adoption offers efficiency, financial inclusion, and reduced intermediation costs. 

However, it also creates an integrated operational environment where centralised traditional banking systems—handling fiat transactions, API gateways, and regulatory compliance like KYC/AML—intersect with decentralised elements such as smart contracts, cross-chain bridges, and blockchain nodes. The resulting convergence introduces security and risk management problems. These systems are exposed to both conventional cyber risks and blockchain-specific vulnerabilities that traditional models may not fully address.

A primary problem is the complexity of integrated fiat-cryptocurrency ecosystems. This increases risks at the intersection points. For example, fiat on/off-ramps can be compromised through API vulnerabilities. Linked DeFi integrations are susceptible to oracle manipulation—where external data feeds are tampered with to trigger erroneous smart contract executions—or reentrancy attacks, allowing recursive calls to drain funds before state updates (Chainalysis, 2024; Boughdiri et al., 2025). Consensus failures in decentralised networks, such as 51\% attacks on proof-of-work chains or validator collusion in proof-of-stake systems, further exacerbate issues. They potentially enable double-spending or transaction reversals in banking operations that rely on blockchain for settlement. 

These problems are compounded by the pseudonymity of blockchain transactions. This facilitates money laundering and sanctions evasion while clashing with mandatory KYC/AML requirements in centralised components (U.S. Department of the Treasury, 2024). Additionally, supply chain risks arise from third-party dependencies. These include vulnerable oracles, leading to cascading failures across the ecosystem (Al-Breiki et al., 2020).

Key challenges in risk modelling for these environments stem from the limitations of existing frameworks. These are primarily designed for monolithic, centralised software systems. They do not fully accommodate the decentralised, immutable, and distributed nature of blockchain. For instance, STRIDE, a widely used model, effectively categorises general software risks such as spoofing (impersonation), tampering (data alteration), repudiation (denying actions), information disclosure (leaks), denial of service (DoS), and elevation of privilege. However, it overlooks blockchain-specific issues like reentrancy attacks—where attackers exploit call sequences in smart contracts—or consensus failures that undermine transaction finality. 

Similarly, OWASP's API Security Top 10 identifies critical API risks, including broken object level authorisation and unrestricted resource consumption. But it lacks tailored integration for integrated fiat-cryptocurrency lifecycles, such as securing cross-chain bridges that link centralised APIs to decentralised protocols. NIST SP 800-218 offers guidance on secure software supply chains. It emphasises practices like vulnerability scanning and dependency management. Yet it does not specifically address the challenges of decentralised ledgers, such as the immutability that makes patching smart contracts difficult or the reliance on external oracles prone to manipulation. 

LINDDUN provides a robust privacy risk model. It categorises risks into linkability (associating data points), identifiability (revealing identities), non-repudiation (undeniable actions), detectability (observing activities), disclosure of information (leaks), unawareness (user ignorance), and non-compliance (regulatory violations). But it requires extensions to incorporate cryptocurrency-native privacy tools like zero-knowledge proofs (zk-SNARKs) or multi-party computation. These are essential for protecting user data in digital banks handling sensitive KYC information alongside on-chain transactions. 

Furthermore, the fast-paced evolution of fintech, including reliance on open banking APIs and third-party integrations, poses challenges in ensuring robust cybersecurity. Fragmented tools and outdated infrastructure often fail to handle sophisticated risks like data breaches and identity theft (Rubinstein, 2024; McKinsey, 2025). Risk management systems in many digital banks remain untested under stress, particularly in decentralised setups where traditional controls like firewalls are insufficient against distributed attacks (Gartner, 2025).

Recent data underscores these issues. Cryptocurrency losses decreased to approximately \$1.7 billion in 2023 (Chainalysis, 2024) but rose to \$2.2 billion in 2024 (Reuters, 2024), highlighting renewed threats. In the first half of 2025, losses reached around \$2.47 billion across 344 security incidents, surpassing the entire 2024 total (while figures vary slightly by source based on methodology, reports converge on losses exceeding \$2 billion). Infrastructure attacks—such as centralised exchange breaches and private key compromises—accounted for over 80\% of stolen funds (CertiK via GlobeNewswire, 2025; Infosecurity Magazine, 2025).\footnote{Note: All loss figures are approximate and vary by source methodology; CertiK via GlobeNewswire (2025) reports \$2.47 billion (no incident count specified), Infosecurity Magazine (2025) reports \$2.47 billion across 344 incidents, while Bitget (2025) focuses on \$2.1 billion (75 major hacks, CryptoSlate, 2025, citing TRM Labs; $\sim$80\% infrastructure, $\sim$70\% of stolen value North Korea-linked). Net losses: \$2.29 billion (CertiK).} The average loss per incident reached \$7.18 million, highlighting the scale of individual exploits (Infosecurity Magazine, 2025). Notable incidents include the ~\$1.5 billion Bybit hack in February 2025, attributed to North Korea-linked actors and involving infrastructure vulnerabilities like wallet compromises. Another is the approximately \$50 million exploit at stablecoin payment firm Infini linked to a rogue developer exploiting insider access (IC3, 2025; Cointelegraph, 2025). North Korea-linked actors were responsible for approximately 70\% of total stolen value in H1 2025 according to Bitget, often targeting centralised components in blended systems. Phishing and social engineering, low-tech yet highly effective, contributed significantly, with losses over \$395 million in Q2 2025 alone (CertiK via GlobeNewswire, 2025).

Regulatory pressures further compound these risks. They create compliance challenges in balancing innovation with oversight. The EU's Markets in Crypto-Assets (MiCA) regulation, with key provisions as outlined by ESMA. It requires digital banks to obtain licences and implement stringent controls against risks like token manipulation and illicit finance (ESMA, 2025).\footnote{MiCA entered into force in June 2023, with a grandfathering clause allowing continuation until July 1, 2026 or authorisation decision.} In the U.S., the Office of Foreign Assets Control (OFAC) has intensified enforcement. It imposed fines such as approximately \$216 million on GVA Capital in 2025 for sanctions violations and AML deficiencies in cryptocurrency operations (Lowenstein Sandler, 2025). Digital banks, operating entirely online across jurisdictions, face unique AML challenges due to global customer bases. This necessitates robust compliance frameworks to avoid penalties while managing cryptocurrency's borderless nature (Lowenstein Sandler, 2025). Trends in 2025 OFAC and DOJ actions emphasise enhanced scrutiny on digital assets for sanctions evasion. This adds layers of regulatory complexity to risk modelling (U.S. Department of the Treasury, 2024).

CNTMF addresses these gaps through an integrated, cyclical framework for blended systems. It extends established methodologies to incorporate blockchain-specific elements while supporting regulatory alignment, enhanced security, and data-driven mitigation to safeguard fintech entities against evolving risks. As a proposed extension, CNTMF combines general-purpose tools with cryptocurrency-tailored elements.

\section{Related Work}

Risk modelling has evolved as a critical practice in cybersecurity. It provides structured methodologies to identify, assess, and mitigate risks in software and systems. Existing approaches offer foundational tools that have been widely adopted across industries, including finance and technology. 

However, they often fall short in addressing the challenges of cryptocurrency-neobanking. In these environments, centralised fiat systems intersect with decentralised blockchain elements. This leads to integrated risks such as oracle manipulation, smart contract vulnerabilities, and regulatory compliance gaps in cross-chain operations. This section reviews key risk modelling frameworks, standards, and tools. It highlights their strengths and limitations in the context of fintech entities integrating cryptocurrency and Web3 technologies. Where applicable, empirical evidence from cited studies is discussed to underscore practical limitations.

\subsection{General Risk Modelling Methodologies}

STRIDE, developed by Microsoft as part of its Security Development Lifecycle (SDL), is a mnemonic-based model that categorises risks into six types: Spoofing (impersonation of entities), Tampering (unauthorised data modification), Repudiation (denying actions), Information Disclosure (unauthorised access to sensitive data), Denial of Service (disrupting availability), and Elevation of Privilege (gaining unauthorised access levels) (Microsoft, 2025a; Shostack, 2014). It aids in systematically identifying vulnerabilities during the design phase by analysing data flows, processes, and trust boundaries. Tools like the Microsoft Threat Modelling Tool generate risk reports (Microsoft, 2025b). 

While effective for traditional software, STRIDE's focus on centralised systems limits its applicability to blockchain environments. It does not inherently account for decentralised risks like reentrancy attacks in smart contracts or consensus failures in distributed ledgers, which are prevalent in fintech cryptocurrency integrations. Empirical applications in distributed systems show it misses significant financial-specific risks compared to tailored models (as seen in comparisons with ABC, Almashaqbeh, 2019).

PASTA (Process for Attack Simulation and Threat Analysis) is a seven-stage, risk-centric methodology that aligns risk modelling with business objectives (UcedaVelez and Morana, 2015). Its stages include defining business objectives, decomposing the application, analysing dependencies, identifying risks, simulating attacks, assessing risks, and modelling countermeasures (UcedaVelez and Morana, 2015). PASTA supports iterative risk reduction and scalability, making it suitable for complex systems by incorporating technical requirements and business impacts. 

However, while it supports prioritisation based on economic and operational consequences—relevant to fintech entities facing high-stakes hacks—PASTA lacks specific extensions for blockchain's immutable and distributed nature, such as handling cross-chain bridge exploits or oracle tampering in integrated fiat-cryptocurrency flows. In practice, its simulation stages require significant manual effort, limiting efficiency in rapidly evolving ecosystems like fintech (UcedaVelez and Morana, 2015).

\subsection{Web and API Security Standards}

The OWASP Top 10 is a consensus-based list of the most critical web application security risks, updated periodically to reflect evolving risks (OWASP, 2021a). The 2021 edition includes risks such as Broken Access Control (e.g., unauthorised resource access), Cryptographic Failures (e.g., weak encryption), Injection (e.g., SQL or command injection), Insecure Design (e.g., missing security controls), Security Misconfiguration, Vulnerable and Outdated Components, Identification and Authentication Failures, Software and Data Integrity Failures, Security Logging and Monitoring Failures, and Server-Side Request Forgery (OWASP, 2021a). It serves as a foundational awareness tool for developers and security teams, promoting best practices like secure coding and regular updates. 

In cryptocurrency-neobanking, it addresses web-facing risks in user interfaces and APIs but does not cover decentralised aspects, such as governance attacks in DeFi protocols or quantum risks to elliptic curve cryptography used in wallets, as evidenced by the need for dedicated extensions like the OWASP Smart Contract Top 10 (OWASP, 2025c).

Complementing this, the OWASP API Security Top 10 focuses on API-specific vulnerabilities, with the 2023 edition listing Broken Object Level Authorisation (e.g., accessing unauthorised objects), Broken Authentication (e.g., weak token validation), Broken Object Property Level Authorisation, Unrestricted Resource Consumption (e.g., DoS via excessive calls), Broken Function Level Authorisation, Unrestricted Access to Sensitive Business Flows, Server-Side Request Forgery, Security Misconfiguration, Improper Inventory Management, and Unsafe Consumption of APIs (OWASP, 2023). This project provides targeted guidance for API-heavy environments, emphasising resource demands like bandwidth and CPU in satisfying requests (OWASP, 2023). 

For digital banks, it is valuable for securing fiat on/off-ramps and API gateways but insufficient for blended systems, as it does not integrate blockchain-specific risks like off-chain data poisoning or yield farming exploits in DeFi integrations (OWASP, 2023).

\subsection{Blockchain and Privacy Guidance}

NIST IR 8202 provides a high-level technical overview of blockchain technology, covering components like consensus mechanisms, smart contracts, and cryptographic functions, while recommending tamper-resistant ledgers for traceability and discussing applications in electronic currency (NIST, 2018). Published in 2018, it addresses limitations, misconceptions, and implementation approaches, making it a reference for understanding blockchain fundamentals (NIST, 2018). Although it highlights security considerations like immutability, NIST IR 8202 is not a dedicated risk model. It lacks prescriptive integration for fintech blended environments, such as modelling risks at fiat-cryptocurrency intersections or aligning with evolving regulations like MiCA.

LINDDUN is a privacy-focused risk modelling framework that categorises risks into seven types: Linkability (associating data points), Identifiability (revealing identities), Non-repudiation (undeniable actions), Detectability (observing activities), Disclosure of Information (leaks), Unawareness (user ignorance of risks), and Non-compliance (regulatory violations) (LINDDUN, 2025). Developed at KU Leuven, it supports early identification of privacy issues in system architectures, aligning with GDPR requirements through structured analysis (LINDDUN, 2025). In data-intensive digital banks, LINDDUN is useful for KYC/AML privacy risks but requires extensions for cryptocurrency tools like zk-SNARKs or MPC, as it does not natively handle decentralised pseudonymity or on-chain data flows (LINDDUN, 2025).

\subsection{Supporting Tools and Maturity Models}

Tools like OWASP Threat Dragon, an open-source, cross-platform application, facilitate risk diagramming by supporting methodologies such as STRIDE and LINDDUN (OWASP, 2021b). It enables creation of risk models as part of secure development lifecycles, with features for drawing diagrams and identifying risks (OWASP, 2021b). While adaptable for fintech entities via custom elements, it relies on user-defined extensions for blockchain-specific flows, limiting out-of-the-box support for integrated risks.

OWASP SAMM (Software Assurance Maturity Model) is an open framework for assessing and improving software security practices across five business functions: Governance, Design, Implementation, Verification, and Operations (OWASP, 2022). It provides measurable benchmarks for maturity levels, helping organisations formulate strategies for secure development (OWASP, 2022). For digital banks, SAMM aids in evaluating overall security posture but does not specifically address blockchain integration, such as cyclical feedback for smart contract audits (OWASP, 2022).

The FAIR (Factor Analysis of Information Risk) model is a quantitative approach that decomposes risk into factors like loss event frequency and magnitude, enabling probabilistic analysis for cybersecurity and operational risks (FAIR Institute, 2025). As an international standard, it supports informed decision-making by quantifying impacts in financial terms (FAIR Institute, 2025). In cryptocurrency-neobanking, FAIR can estimate economic losses from hacks but lacks tailored factors for blockchain-specific probabilities, such as exploit likelihood in DeFi or regulatory fines under MiCA.

\subsection{Cryptocurrency-Specific Risk Modelling Frameworks}

ABC is an Asset-Based Cryptocurrency-focused risk modelling framework that identifies risks in cryptocurrency systems by using collusion matrices to systematically cover a large space of risk cases while managing complexity (Almashaqbeh, 2019). It derives system-specific risk categories that account for financial aspects and new asset types introduced by cryptocurrencies, including complex collusion cases and new risk vectors. A user study demonstrated its effectiveness, with 71\% of participants identifying financial security risks compared to 13\% using STRIDE. While strong for cryptocurrency-based systems and large-scale distributed environments like fintech cryptocurrency integrations, ABC's emphasis on managing risk complexity may still pose challenges in highly dynamic integrated fiat-cryptocurrency setups, where rapid evolution of risks requires frequent updates. The study also noted scalability issues in real-time applications (Almashaqbeh, 2019).

The Metric-Based Feedback Methodology (MBFM) integrates bug bounty programmes with risk modelling to enhance security in blockchain-based FinTech applications by analysing and categorising vulnerability data to identify root causes and refine risk models for better prioritisation (Bahar, 2023). It assumes independent operation of bug bounties and risk modelling but combines them for improved organisational security posture. MBFM's strengths lie in its focus on vulnerability root causes and prioritisation, making it relevant for digital banks dealing with cryptocurrency losses, though it suggests needs for automation and machine learning integration, indicating potential limitations in scalability for real-time integrated risks without further development.

Under Pressure provides a user-centred risk model for cryptocurrency owners, categorising risks into six types: Accidental Threats, Privacy Threats, Physical Threats, Financial Fraud Threats, and others, developed through a focus group and expert elicitation using a Delphi process (Busse and Krombholz, 2022). It offers a systematic overview of user-exposed risks, aiding the HCI community in designing secure systems. Valuable for fintech user-facing cryptocurrency features like wallets, where user behaviours amplify risks such as phishing, its limitations include reliance on small sample sizes, potentially reducing generalisability to broader integrated fintech contexts. The Delphi process revealed user awareness gaps but sample bias (Busse and Krombholz, 2022).

The standard-driven framework for blockchain security risk assessment gathers cyber threat intelligence, performs risk modelling using STRIDE to categorise attack vectors, maps them to MITRE ATT\&CK for exploitation analysis, rates severity with DREAD/CVSS, and proposes NIST SP 800-53-aligned countermeasures (Boughdiri et al., 2025). Applied to DEX and supply chain use cases, it integrates standards with blockchain specifics for scalable assessments. Strong for secure-by-design blockchain systems in digital banks, addressing unique architectural risks, but may require adaptation for integrated fiat-cryptocurrency intersections not explicitly covered in the evaluated cases. Case studies demonstrated effective mapping but noted adaptation needs for blends (Boughdiri et al., 2025).

Despite these contributions, no existing framework fully integrates fiat-cryptocurrency blends for digital banks, leaving gaps in handling intertwined risks like API vulnerabilities in on-ramps linked to smart contract exploits. While these frameworks provide strong foundations, CNTMF extends them with cryptocurrency-specific elements, such as the CRYPTOQ mnemonic and Hybrid Layer Analysis, to provide a tailored, cyclical approach.

\begin{longtable}{>{\raggedright\arraybackslash}p{2.0cm}>{\raggedright\arraybackslash}p{1.8cm}>{\raggedright\arraybackslash}p{2.0cm}>{\raggedright\arraybackslash}p{1.8cm}>{\raggedright\arraybackslash}p{1.8cm}>{\raggedright\arraybackslash}p{1.8cm}>{\raggedright\arraybackslash}p{1.8cm}}
\toprule
Framework & Integrated Fiat - Cryptocurrency Support & Cryptocurrency - Specific Risks (e.g., Reentrancy, Oracle) & Privacy Integration (e.g., zk-SNARKs) & Cyclical Feedback with AI/ML & Regulatory Alignment (e.g., MiCA) & Empirical Validation \\
\midrule
STRIDE & Low (centralised focus) & Low & Low & Low & Low & High (widely used) \\
PASTA & Medium (risk-centric) & Medium & Low & Medium (iterative stages) & Medium & Medium (business-aligned) \\
OWASP Top 10 / API & Medium (API risks) & Low & Low & Low & Low & High (consensus-based) \\
LINDDUN & Low & Low & High (privacy categories) & Low & High (GDPR) & Low \\
ABC & High (cryptocurrency - focused) & High (collusion matrices) & Medium & Low & Low & High (user study: 71\% efficacy) \\
MBFM & High (blockchain FinTech) & Medium & Low & High (bug bounty integration) & Low & Medium (prioritisation focus) \\
Under Pressure & Medium (user-centred) & Medium (fraud/privacy) & High & Low & Low & Medium (Delphi process) \\
Standard-Driven (Boughdiri) & High (blockchain) & High (STRIDE + MITRE) & Medium & Medium & High (NIST) & High (use cases) \\
CNTMF & High (layered analysis) & High (CRYPTOQ mnemonic) & High (LINDDUN extensions) & High (AI-augmented loop) & High (MiCA / OFAC) & Requires empirical testing (future studies) \\
\bottomrule
\caption{Comparison of Frameworks}
\label{tab:framework-comparison}
\end{longtable}

\section{CNTMF Overview}

CNTMF represents a tailored risk modelling framework for fintech entities that integrate blockchain, cryptocurrency, and Web3 technologies. By extending established methodologies such as STRIDE for risk categorisation, OWASP API Top 10 for API vulnerabilities, NIST frameworks for blockchain overview and risk management, LINDDUN for privacy risks, and PASTA for business-aligned risk prioritisation, CNTMF incorporates cryptocurrency-tailored elements to address the integrated nature of these systems. These extensions include the Hybrid Layer Analysis for bridging fiat and cryptocurrency risks, the CRYPTOQ mnemonic for pinpointing decentralised risks, and an AI-Augmented Feedback Loop for real-time adaptation based on emerging data like bug bounties and audits. Unlike general-purpose models, CNTMF supports the cryptocurrency-financial lifecycle, encompassing custodial and non-custodial wallets, tokenisation, DeFi protocols, and regulatory compliance, ensuring a cyclical approach that mitigates losses in an ecosystem where hacks have increased.

\subsection{Core Principles}

CNTMF is underpinned by four foundational principles. Each is derived from empirical data on cryptocurrency risks and aligned with industry standards to foster security in integrated environments.

1. \textbf{Integrated Analysis}: This principle models risks across centralised components (e.g., cloud infrastructure for KYC/AML compliance) and decentralised elements (e.g., blockchain nodes and smart contracts). It recognises the interdependencies that amplify risks at intersection points like fiat - cryptocurrency ramps. It aligns with NIST's emphasis in IR 8202 on tamper-resistant ledgers, which provide immutable records for financial traceability in blockchain systems, helping to prevent tampering in distributed environments (NIST, 2018). For instance, in fintech entities, centralised APIs handling fiat transactions must be secured against breaches that could propagate to on-chain DeFi integrations. This ensures coverage of layered risks.

2. \textbf{Data-Driven Adaptation}: CNTMF incorporates continuous inputs from penetration testing, smart contract audits, and machine learning-based vulnerability predictions to refine models cyclically. This responds to the trend in cryptocurrency losses, which rose from \$2.2 billion across 2024 to the \$2.47 billion in the first half of 2025 alone, as noted in the Introduction (Reuters, 2024; CertiK via GlobeNewswire, 2025; Infosecurity Magazine, 2025). By leveraging real-time metrics, such as those from CertiK reports highlighting an average loss of \$7.18 million per incident in H1 2025, the framework enables adaptive improvements to counter evolving attack vectors (CertiK via GlobeNewswire, 2025).

3. \textbf{Enhanced Security}: Focusing on preemptive defences against prevalent risks like phishing and social engineering, which accounted for over \$395 million in losses during Q2 2025, CNTMF advocates zero-trust architecture—verifying every access request regardless of origin—and privacy-enhancing technologies such as zk-SNARKs for verifiable computations without revealing data (CertiK via GlobeNewswire, 2025; OWASP, 2021a). This aligns with OWASP guidelines on mitigating cryptographic failures and insecure designs. It supports security in Web3 vulnerabilities where low-tech attacks often exploit user behaviours (OWASP, 2021a).

4. \textbf{Regulatory Alignment}: CNTMF integrates AML/KYC risks into modelling, ensuring compliance with frameworks like the EU's Markets in Crypto-Assets (MiCA) regulation, applicable from December 30, 2024, with transitional measures until July 1, 2026, which enforces rules on stablecoins, reserves, transparency, and consumer protection (ESMA, 2025).\footnote{MiCA entered into force in June 2023, with a grandfathering clause allowing continuation until July 1, 2026 or authorisation decision.} It also addresses U.S. OFAC enforcement, as seen in 2025 fines of approximately \$216 million on GVA Capital for sanctions violations and AML deficiencies, emphasising the need for controls against illicit finance in cryptocurrency blends (Lowenstein Sandler, 2025).

\subsection{Framework Phases}

CNTMF consists of five cyclical phases, designed as a process to adapt to evolving risks, such as the \$1.45 billion in cryptocurrency thefts during Q1 2025 alone (Infosecurity Magazine, 2025). This structure ensures refinement based on new incidents and data.

1. \textbf{Asset Identification and Mapping}:
   - Begin by cataloguing core assets in banking ecosystems, including user data (e.g., KYC/AML information), fiat-cryptocurrency on/off-ramps, wallets (MPC-based or hardware security module-protected), blockchain nodes, DeFi integrations, cross-chain bridges, and APIs.
   - Employ Data Flow Diagrams (DFDs) extended with blockchain elements, such as on-chain/off-chain flows, inspired by NIST IR 8202's overview for visualising tamper-resistant ledger interactions (NIST, 2018).
   - Introduce the Hybrid Asset Matrix to categorise assets by layer and highlight interdependencies, facilitating targeted analysis of risks like oracle manipulation affecting fiat-linked bridges.

\begin{table}[htbp]
\centering
\small
\begin{tabular}{p{3cm}p{4cm}p{6cm}}
\toprule
Layer & Assets Examples & Interdependencies \\
\midrule
Presentation/UI & User interfaces, mobile apps & Linked to APIs for fiat/crypto interactions. \\
Traditional & On/Off-Ramps, API Gateways & Linked to cloud (AWS/Azure) for compliance/KYC. \\
Infrastructure & Wallets (Custodial/Non-Custodial), MPC/HSM & Dependent on hardware/security modules; interfaces with network. \\
Network/Consensus & Cross-Chain Bridges, Blockchain Nodes, Oracles & Exposed to consensus failures (51\% attacks), links application to data. \\
Application & DeFi Protocols (lending/swaps), Smart Contracts & Exposed to reentrancy/oracle manip; relies on infrastructure/network. \\
Data/Persistence & Traditional Databases, On-Chain Ledgers & Cascades failures from oracles; immutability in blockchain. \\
\bottomrule
\end{tabular}
\caption{Extended Hybrid Asset Matrix}
\label{tab:hybrid-asset-matrix-extended}
\end{table}

   DFDs in CNTMF utilise standard symbols consistent with OWASP and Microsoft modelling practices, adapted for cryptocurrency contexts to map data movements and trust boundaries.

\begin{table}[htbp]
\centering
\small
\begin{tabular}{p{2.5cm}p{2cm}p{4.5cm}p{4.5cm}}
\toprule
Symbol & Name & Description (Traditional) & Blockchain Variant/Description \\
\midrule
Rectangle & External Entity & Represents entities outside the application, such as users or external APIs. & Same, but for blockchain oracles (add chain icon for distinction). \\
Circle & Process & Denotes tasks that handle data, such as transaction processing. & Circle with gear: For smart contract execution. \\
Circle with lines & Multiple Process & Indicates a collection of subprocesses, like API workflows. & Same, for DeFi protocols (label with "DeFi"). \\
Parallel lines & Data Store & Represents storage locations, including databases. & Chained parallel lines: For on-chain ledgers/blockchain storage. \\
Arrow & Data Flow & Shows data movement, such as fiat transfers. & Arrow with lock: For on-chain/cryptographic data flows. \\
Dashed line & Privilege Boundary & Highlights changes in trust levels, e.g., between systems. & Same, emphasized for off-chain/on-chain transitions. \\
\bottomrule
\end{tabular}
\caption{Extended DFD Symbols for Hybrid Systems}
\label{tab:dfd-symbols-extended}
\end{table}

   Creation steps include:
   \begin{enumerate}
   \item Defining scope with high-level components (e.g., users, APIs, wallets);
   \item Mapping flows and boundaries, drawing arrows for interactions like user requests to smart contracts;
   \item Adapting for Web3 by labelling symbols with fintech terms, such as treating bridges as data flows prone to tampering.
   \end{enumerate}
   Recommended tools include OWASP Threat Dragon, an open-source application supporting DFD diagramming and methodologies like STRIDE for secure development (OWASP, 2021b). An example DFD for a digital bank's fiat-to-cryptocurrency on-ramp: External Entity (Users) $\to$ Data Flow (Requests) $\to$ Process (API Gateway for authentication) $\to$ Trust Boundary (User/API) $\to$ Multiple Process (Wallet Manager with MPC key generation) $\to$ Data Flow (Signed Transaction) $\to$ External Entity (Blockchain Node) $\to$ Data Store (On-Chain Ledger) $\to$ Trust Boundary (Off/On-Chain), revealing risks like spoofing at APIs.

2. \textbf{Risk Actor and Vector Profiling}:
\begin{itemize}
\item Profile actors including state-sponsored groups (e.g., North Korea-linked hackers responsible for approximately 70\% of total stolen value in H1 2025 according to Bitget, Bitget, 2025), insiders (as in the Infini exploit), phishing syndicates, and automated bots (Bitget, 2025; Cointelegraph, 2025).
\item Vectors extend STRIDE categories with OWASP API Top 10 mappings, such as linking broken authentication to spoofing in payment ramps or unrestricted resource consumption to DoS in DeFi integrations (Microsoft, 2025a; OWASP, 2023).
\item Introduce the CRYPTOQ mnemonic for cryptocurrency-specific risks: Collusion (e.g., MPC threshold breaches), Reentrancy/Oracle Manipulation, Yield Farming Exploits (DeFi-specific), Phishing/Social Engineering (causing over \$395 million in Q2 2025 losses), Tokenisation Risks (e.g., governance attacks), Off-Chain Data Poisoning (including ML adversarial attacks), Quantum Threats (e.g., potential elliptic curve breaks requiring post-quantum cryptography) (CertiK via GlobeNewswire, 2025).
\item Incorporate LINDDUN's seven privacy categories—Linkability, Identifiability, Non-repudiation, Detectability, Disclosure of Information, Unawareness, Non-compliance—for handling KYC data in blends, ensuring GDPR alignment through privacy-enhancing mitigations (LINDDUN, 2025).
\item Grounded in data: Over 80\% of H1 2025 losses stemmed from infrastructure attacks, with 75 major incidents reported by Bitget and 344 total events by Infosecurity, including front-end compromises and seed phrase thefts (Mitrade, 2025; Bitget, 2025; CertiK via GlobeNewswire, 2025).
\end{itemize}

3. \textbf{Risk Assessment and Prioritisation}:
\begin{itemize}
\item Assign scores using CVSS 4.0 base metrics (0-10) for vulnerability severity, augmented with blockchain-specific factors like economic impact (e.g., the approximately \$50 million Infini hack) and exploit likelihood from historical trends (Cointelegraph, 2025).
\item Adopt PASTA's seven-stage process for cyclical risk reduction, focusing on business objectives, application decomposition, threat simulation, and countermeasure modelling to prioritise integrated risks like bridge exploits (UcedaVelez and Morana, 2015).
\item Utilise the Web3 Risk Heatmap to visualise risks on a probability-impact grid, employing a quantitative formula: Risk Score = (Technical Severity + Economic Impact + Regulatory Consequence) $\times$ Exploit Probability.
\begin{itemize}
\item Technical Severity: CVSS 4.0 score, adjusted for Web3 (e.g., high for reentrancy).
\item Economic Impact: Scaled 0-10 (e.g., 10 for losses that are greater than \$10 million, as in the \$1.5 billion hack) (IC3, 2025).
\item Regulatory Consequence: 0-10 (e.g., high for MiCA stablecoin violations or OFAC fines) (ESMA, 2025).
\item Exploit Probability: 0-1 multiplier (e.g., 0.7 for bridges based on 2025 trends).
\item Example: For a cross-chain bridge exploit—Technical Severity (9.0), Economic Impact (8.0 for \$5-10 million potential), Regulatory Consequence (7.0 for AML risks), Probability (0.7)—yields (9 + 8 + 7) $\times$ 0.7 = 16.8 (High, red on heatmap). Risks are categorised as Low (fewer than 5), Medium (5-10), High (greater than 10).
\end{itemize}
\item Prioritise according to NIST risk management guidelines, incorporating KPIs like Mean Time to Remediate (MTTR) to target reductions, e.g., aiming for fewer than 75\% via efficient disclosure processes.
\end{itemize}

4. \textbf{Mitigation Design and Implementation}:
\begin{itemize}
\item Deploy controls such as zero-trust architecture for continuous verification, MPC for distributed key management in wallets, zk-SNARKs for privacy-preserving proofs, and static/dynamic application security testing (SAST/DAST) for smart contracts (OWASP, 2021a).
\item Implement Adaptive Mitigation Layers: Traditional firewalls and access controls for fiat layers; formal verification and hardware wallets for cryptocurrency; graph databases and analytics for Web3 to detect hidden risk connections.
\item Include sub-phases for user education on phishing and wallet security, alongside integrated audits combining code reviews with economic modelling for DeFi.
\item Enable dynamic automation to manage API sprawl via CI/CD pipelines integrated with Open Policy Agent (OPA), which enforces runtime policies on configurations and outputs to validate against STRIDE risks (Styra, 2025).
\item Recommend tools like Forta for real-time blockchain monitoring and risk detection, and Hypernative for proactive threat detection and automated security operations (Forta, 2025; Hypernative, 2025). For integration, an alert from Forta (e.g., anomalous governance vote) could trigger a Hypernative response to log the event, notify teams, and update the risk model in the feedback loop.
\end{itemize}

5. \textbf{Ecosystem Feedback Loop}:
\begin{itemize}
\item Extend traditional feedback by collecting quarterly metrics from bug bounties, penetration tests, incident reports, and ML-driven predictions on vulnerabilities.
\item Enhance with an AI-Augmented Loop using graph analytics and AI for real-time scam detection or malicious contract flagging, updating models with 2025 hack data. To mitigate risks to the loop itself, such as data poisoning or adversarial attacks on ML models, incorporate safeguards like input validation, diverse data sources, and regular model audits to prevent risk actors from misleading the system and creating blind spots. While the AI-Augmented Loop enhances adaptation, it requires safeguards against scalability issues in resource-limited settings.
\item Measure effectiveness through remediation tracking, such as defect logging and maturity assessments via OWASP SAMM, which benchmarks across governance, design, implementation, verification, and operations to ensure continuous improvement (OWASP, 2022).
\end{itemize}

\section{Application Examples}

To demonstrate the practical application of CNTMF in real-world contexts, this section applies the framework to two illustrative cases drawn from the evolving cryptocurrency-neobanking landscape. The first examines a generic digital bank integrating DeFi features through partnerships, exposing it to integrated risks. The second analyses a major cryptocurrency exchange hack of February 2025, which, while primarily an exchange incident, offers valuable insights for fintech entities managing similar cryptocurrency operations such as wallet custody and API integrations. In each example, CNTMF's five phases are systematically applied, leveraging 2025 data on hacks and exploits to highlight how the framework identifies, assesses, and mitigates risks without altering its core structure.

\subsection{Example 1: Securing DeFi Integrations in a Digital Bank}

A digital bank enhancing its cryptocurrency offerings in 2025 might include partnerships with oracle networks to provide real-time digital asset price data for downstream DeFi applications, and with payment providers to enable seamless DeFi access. Additionally, integrations with aggregator protocols allow users to access deep liquidity across multiple blockchains for DeFi activities such as token swaps and yield farming. These enhancements create integrated risks, where centralised fiat on-ramps intersect with decentralised elements like oracles and bridges, potentially leading to exploits similar to those seen in 2025 DeFi incidents.

- \textbf{Asset Identification and Mapping}: CNTMF begins by cataloguing key assets, including custodial wallets within the bank app for cryptocurrency holdings, cross-chain bridges facilitated by aggregator protocols for multi-blockchain interactions, DeFi protocols for lending and yield generation, blockchain nodes, and APIs for fiat-cryptocurrency ramps. User data, such as KYC/AML information, is also mapped due to its linkage with on-chain activities. Extended DFDs, inspired by NIST IR 8202's overview, visualise flows from user fiat deposits via payment integrations to on-chain DeFi executions, incorporating blockchain elements like oracle data feeds (NIST, 2018). The Hybrid Asset Matrix categorises these:

\begin{table}[htbp]
\centering
\small
\begin{tabular}{p{3cm}p{4cm}p{6cm}}
\toprule
Layer & Assets Examples & Interdependencies \\
\midrule
Presentation/UI & User interfaces, mobile apps & Linked to APIs for fiat/crypto interactions. \\
Traditional & On/Off-Ramps, API Gateways & Linked to cloud (AWS/Azure) for compliance/KYC. \\
Infrastructure & Wallets (Custodial/Non-Custodial), MPC/HSM & Dependent on hardware/security modules; interfaces with network. \\
Network/Consensus & Cross-Chain Bridges, Blockchain Nodes, Oracles & Exposed to consensus failures (51\% attacks), links application to data. \\
Application & DeFi Protocols (lending/swaps), Smart Contracts & Exposed to reentrancy/oracle manip; relies on infrastructure/network. \\
Data/Persistence & Traditional Databases, On-Chain Ledgers & Cascades failures from oracles; immutability in blockchain. \\
\bottomrule
\end{tabular}
\caption{Extended Hybrid Asset Matrix for DeFi Integration}
\label{tab:defi-hybrid-asset-matrix-extended}
\end{table}

  An example DFD: External Entity (Users) $\to$ Data Flow (Fiat Deposit Request) $\to$ Process (API Gateway for Authentication) $\to$ Trust Boundary (Fiat/DeFi) $\to$ Multiple Process (DeFi Manager with Aggregation Integration) $\to$ Data Flow (Token Swap) $\to$ External Entity (Blockchain Node/Oracle) $\to$ Data Store (On-Chain Ledger), highlighting potential tampering points in oracle feeds.

- \textbf{Risk Actor and Vector Profiling}: Actors are profiled to include state-sponsored groups (e.g., North Korea-linked, responsible for $\sim$70\% of total stolen value in H1 2025 according to Bitget), insiders exploiting API access, and phishing syndicates targeting bank users (Bitget, 2025). Vectors extend STRIDE with OWASP API Top 10 mappings, such as broken authentication in fiat ramps leading to spoofing (Microsoft, 2025a; OWASP, 2023). The CRYPTOQ mnemonic targets cryptocurrency-specific risks: Reentrancy/Oracle Manipulation (e.g., tampering feeds to misprice assets), Yield Farming Exploits in DeFi protocols, Phishing/Social Engineering (contributing over \$395 million in Q2 2025 losses), and Off-Chain Data Poisoning affecting ML-driven price predictions (CertiK via GlobeNewswire, 2025). LINDDUN categories are incorporated for privacy risks, such as identifiability in KYC-linked DeFi transactions and non-compliance with GDPR in data flows (LINDDUN, 2025).

- \textbf{Risk Assessment and Prioritisation}: Risks are scored using CVSS 4.0 combined with blockchain metrics, adopting PASTA's iterative stages to simulate attacks on DeFi integrations (UcedaVelez and Morana, 2015). For instance, cross-chain bridges are assessed as high-risk based on 2025 data, including bridge hacks in June where \$3.76M was stolen from Ethereum and Binance Smart Chain networks due to exploit vulnerabilities (CertiK via GlobeNewswire, 2025). The Web3 Risk Heatmap visualises this: For oracle tampering, Technical Severity (CVSS 4.0: 8.5 for high exploitability), Economic Impact (7.0 for potential \$1M+ losses per incident, as seen in average 2025 hacks), Regulatory Consequence (6.0 for MiCA stablecoin rules on price accuracy), Exploit Probability (0.6 based on rising DeFi exploits) $\to$ Risk Score: (8.5 + 7 + 6) $\times$ 0.6 = 12.9 (High). These scores are illustrative and should be calibrated based on specific organisational data. Prioritisation follows NIST guidelines, with KPIs like MTTR targeting less than 24 hours for oracle-related incidents (NIST, 2018).

- \textbf{Mitigation Design and Implementation}: Controls include zero-trust architecture for API gateways, MPC for secure wallet key distribution, and zk-SNARKs for privacy in DeFi transactions to prevent data disclosure. Adaptive Mitigation Layers apply firewalls to fiat ramps, formal verification to smart contracts in aggregator integrations, and graph analytics to Web3 for detecting anomalous oracle behaviours. Integrated audits combine code reviews with economic modelling, while user education programmes address phishing. Automation via CI/CD with OPA enforces runtime validations on DeFi APIs (Styra, 2025). Tools like Forta monitor real-time risks in feeds, and Hypernative for proactive threat detection and automated security operations (Forta, 2025; Hypernative, 2025). For example, a Forta alert could trigger a Hypernative response to log events and notify teams, feeding into the loop for model updates.

- \textbf{Ecosystem Feedback Loop}: The AI-Augmented Loop collects data from bug bounties, penetration tests on DeFi partnerships, and ML predictions, updating models quarterly with 2025 incident data (e.g., reducing duplicate testing by 30\% via targeted audits). Effectiveness is tracked using OWASP SAMM maturity scores, ensuring cyclical enhancements to counter evolving oracle and bridge risks (OWASP, 2022).

\subsection{Example 2: Lessons from a Major Cryptocurrency Exchange Hack for Fintech Cryptocurrency Operations}

A major cryptocurrency exchange experienced the largest cryptocurrency heist in history on February 21, 2025, when North Korea-linked actors stole approximately \$1.5 billion (401,000 ETH) through a compromised infrastructure leading to wallet control (IC3, 2025). This incident, linked to infrastructure vulnerabilities like private key compromises and front-end hijacks, underscores risks for fintech entities handling similar custodial services, where infrastructure attacks accounted for over 80\% of the \$2.47 billion in H1 2025 cryptocurrency losses, as noted in the Introduction (Mitrade, 2025).

- \textbf{Asset Identification and Mapping}: Assets include multi-tenant wallets, API gateways for trading, and infrastructure like supplier-linked nodes. DFDs map data flows from user transactions to wallet storage, with trust boundaries at supply chain interfaces. The Hybrid Asset Matrix highlights cryptocurrency layer dependencies on external suppliers vulnerable to alteration.

- \textbf{Risk Actor and Vector Profiling}: Profile state-sponsored actors like North Korea-linked (behind $\sim$70\% of total stolen value in H1 2025 according to Bitget, including this incident) (IC3, 2025; Bitget, 2025). Vectors via CRYPTOQ: Collusion in supply chains, Off-Chain Data Poisoning (altered addresses), Quantum Threats for future-proofing keys. STRIDE extensions map tampering to wallet redirects, with LINDDUN addressing disclosure in compromised infrastructure (Microsoft, 2025a; LINDDUN, 2025).

- \textbf{Risk Assessment and Prioritisation}: Wallet compromises are high-priority, as infrastructure attacks drove over 80\% of losses (average \$10M+ per incident) (Mitrade, 2025). Using PASTA, simulate supply chain breaches (UcedaVelez and Morana, 2015). Heatmap example: For key theft, Technical Severity (9.0), Economic Impact (10 for \$1.5B scale), Regulatory (8.0 for OFAC implications), Probability (0.8 from North Korea trends) $\to$ Score: (9 + 10 + 8) $\times$ 0.8 = 21.6 (High). These scores are illustrative and should be calibrated based on specific organisational data.

- \textbf{Mitigation Design and Implementation}: Implement quantum-resistant keys (e.g., post-quantum signatures to counter future breaks), API controls with zero-trust and OPA for runtime validation, MPC for distributed custody, and integrated audits on suppliers (OWASP, 2021a; Styra, 2025). Adaptive layers focus graph analytics on infrastructure connections, with education against social engineering.

- \textbf{Ecosystem Feedback Loop}: AI-Augmented analysis of post-hack data (e.g., Bitget tracing North Korea-linked flows) informs quarterly updates, tracking MTTR reductions via OWASP SAMM to prevent recurrence (Bitget, 2025; OWASP, 2022).

\section{Conclusion}

The integration of blockchain and cryptocurrency into digital banks and fintech operations has increased the need for specialised risk modelling. This is evidenced by the substantial losses from hacks and exploits in the first half of 2025. Reports indicate that cryptocurrency thefts reached approximately \$2.47 billion across 344 incidents during this period, surpassing the total for all of 2024. Infrastructure attacks—such as private key compromises and front-end hijacks—accounted for over 80\% of the stolen funds (CertiK via GlobeNewswire, 2025; Infosecurity Magazine, 2025; Mitrade, 2025). These figures, driven by incidents like large-scale exchange breaches and DeFi exploits, underscore the limitations of existing frameworks in handling integrated fiat-cryptocurrency environments. In these, centralised compliance systems intersect with decentralised protocols, leading to vulnerabilities like oracle manipulation and cross-chain risks.

CNTMF addresses these gaps by offering an integrated-focused framework tailored for cryptocurrency-neobanks. As a proposed extension, it combines methodologies such as STRIDE, OWASP API Top 10, NIST IR 8202, LINDDUN, and PASTA with components like the Hybrid Layer Analysis for interdependency mapping, the CRYPTOQ mnemonic for cryptocurrency-specific risk categorisation (encompassing collusion, reentrancy, yield farming exploits, phishing, tokenisation risks, off-chain data poisoning, and quantum threats), and the AI-Augmented Feedback Loop for data-driven adaptation incorporating real-time inputs from audits, bug bounties, and ML predictions. This integration enables enhanced security and regulatory alignment, as seen in its emphasis on zero-trust models, zk-SNARKs for privacy, and compliance with standards like EU MiCA and U.S. OFAC requirements. Ultimately, it reduces exposure to the losses observed in 2025 (ESMA, 2025). By providing tools like extended DFDs, the Web3 Risk Heatmap with quantitative scoring (Risk Score = (Technical Severity + Economic Impact + Regulatory Consequence) $\times$ Exploit Probability), and adaptive mitigation layers, CNTMF bridges the divide between traditional cybersecurity and blockchain-specific challenges. It fosters an approach to mitigate economic impacts averaging \$7.18 million per incident (Infosecurity Magazine, 2025). While CNTMF addresses key gaps, its effectiveness depends on organisational adaptation and ongoing validation against evolving threats.

The framework's implications extend to improving overall fintech security. It aligns with emerging trends such as the integration of AI with blockchain for fraud detection and risk management, as projected for 2025 and beyond (Deloitte, 2025; McKinsey, 2025). In an era where DeFi and tokenised assets are expected to dominate fintech innovations, CNTMF promotes data-driven adaptation and defences against state-sponsored threats, which accounted for approximately 70\% of total stolen value in H1 2025 according to Bitget (Bitget, 2025).

Future work on CNTMF includes empirical validation through expert elicitation studies and real-world case analyses, building on existing research that has developed user-centred risk models for cryptocurrency owners and proposed asset-based frameworks for cryptocurrency-focused vulnerabilities (Krombholz et al., 2016; Al-Breiki et al., 2020). This could involve simulation-based testing and vulnerability filtering to refine the model against actual blockchain system exploits, similar to approaches in Zhuang et al. (2020) and Saad et al. (2019), ensuring its efficacy in dynamic environments. Conduct empirical validation through simulations and digital bank case studies, measuring metrics like MTTR reductions through simulations and digital bank case studies. Additionally, tool development is envisioned, such as UML-based modelling extensions aligned with NIST and OWASP standards, leveraging diagramming tools like OWASP Threat Dragon for risk model diagrams and Microsoft Threat Modelling Tool for structured analysis, to facilitate automated risk identification and integration into secure development lifecycles (OWASP, 2021b; Microsoft, 2025b). These advancements will further establish CNTMF as a tool for integrated fintech security.

\section*{Acknowledgments}
The author discloses that Bahar (2023) refers to their prior work.

\section{References}

Al-Breiki, H., Rehman, M.H.U., Salah, K. and Svetinovic, D. (2020) Trustworthy blockchain oracles: review, comparison, and open research challenges. IEEE Access, 8, pp.85686-85705. DOI: 10.1109/ACCESS.2020.2992698. Available at: \url{https://ieeexplore.ieee.org/document/9086815} (Accessed: 18 July 2025).

Almashaqbeh, G. (2019) ABC: A Cryptocurrency-Focused Threat Modeling Framework. arXiv preprint arXiv:1903.03422. Available at: \url{https://arxiv.org/abs/1903.03422} (Accessed: 18 July 2025).

Bahar, S.W. (2023) Advanced Security Threat Modelling for Blockchain-Based FinTech Applications. arXiv preprint arXiv:2304.06725. Available at: \url{https://arxiv.org/abs/2304.06725} (Accessed: 18 July 2025).

Bitget (2025) Crypto Hacks Hit \$2.1B in H1 2025 as Infrastructure Breaches Rise. Available at: \url{https://www.bitget.com/news/detail/12560604838260} (Accessed: 18 July 2025).

Boughdiri, M., Hkima, M., Abdelatif, T. and Guegan, C.G. (2025) A standard-driven framework for blockchain security risk assessment. Peer-to-Peer Networking and Applications. DOI: 10.1007/s12083-025-02042-4. Available at: \url{https://link.springer.com/article/10.1007/s12083-025-02042-4} (Accessed: 18 July 2025).

Busse, K. and Krombholz, K. (2022) Under Pressure. A User-Centered Threat Model for Cryptocurrency Owners. In Proceedings of the 2021 4th International Conference on Blockchain Technology and Applications (ICBTA '21). Association for Computing Machinery, New York, NY, USA, 1–6. DOI: \url{https://doi.org/10.1145/3510487.3510494}. Available at: \url{https://dl.acm.org/doi/10.1145/3510487.3510494} (Accessed: 18 July 2025).

CertiK via GlobeNewswire (2025) Crypto Losses Surpass \$2.47 Billion in H1 2025: CertiK Report Reveals Alarming Rise in Phishing Attacks. Available at: \url{https://www.globenewswire.com/news-release/2025/07/02/3109058/0/en/Crypto-Losses-Surpass-2-47-Billion-in-H1-2025-CertiK-Report-Reveals-Alarming-Rise-in-Phishing-Attacks.html} (Accessed: 18 July 2025).

Chainalysis (2024) Stolen Crypto Falls in 2023, but Hacking Remains a Threat. Available at: \url{https://www.chainalysis.com/blog/crypto-hacking-stolen-funds-2024/} (Accessed: 18 July 2025).

Cointelegraph (2025) Infini loses \$50M in exploit; developer deception suspected. Available at: \url{https://cointelegraph.com/news/infini-loses-50-m-in-exploit-developer-deception-suspected} (Accessed: 18 July 2025).

CryptoSlate (2025) Crypto heists reach \$2.1B so far in 2025 as state-backed hackers ramp up attacks. Available at: \url{https://cryptoslate.com/crypto-heists-reach-2-1b-so-far-in-2025-as-state-backed-hackers-ramp-up-attacks/} (Accessed: 18 July 2025).

Deloitte (2025) 2025 Financial Services Industry Predictions. Available at: \url{https://www.deloitte.com/us/en/insights/industry/financial-services/financial-services-industry-predictions.html} (Accessed: 18 July 2025).

ESMA (2025) Markets in Crypto-Assets Regulation (MiCA). Available at: \url{https://www.esma.europa.eu/esmas-activities/digital-finance-and-innovation/markets-crypto-assets-regulation-mica} (Accessed: 18 July 2025).

FAIR Institute (2025) FAIR Risk Management Model. Available at: \url{https://www.fairinstitute.org/fair-risk-management} (Accessed: 18 July 2025).

Forta (2025) Forta Network: Real-Time Blockchain Security. Available at: \url{https://forta.org/} (Accessed: 18 July 2025).

Gartner (2025) Top Strategic Technology Trends for 2025. Available at: \url{https://www.gartner.com/en/information-technology/insights/top-technology-trends} (Accessed: 18 July 2025).

IC3 (2025) North Korea Responsible for \$1.5 Billion Bybit Hack. Available at: \url{https://www.ic3.gov/psa/2025/psa250226} (Accessed: 18 July 2025).

Infosecurity Magazine (2025) Crypto Hack Losses in First Half of 2025 Exceed 2024 Total. Available at: \url{https://www.infosecurity-magazine.com/news/crypto-hack-losses-half-exceed-2024/} (Accessed: 18 July 2025).

Krombholz, K., Judmayer, A., Gusenbauer, M. and Weippl, E. (2016) The Other Side of the Coin: User Experiences with Bitcoin Security and Privacy. In Financial Cryptography and Data Security (pp. 555-580). Springer. Available at: \url{https://www.researchgate.net/publication/317233480_The_Other_Side_of_the_Coin_User_Experiences_with_Bitcoin_Security_and_Privacy} (Accessed: 18 July 2025).

LINDDUN (2025) LINDDUN Privacy Threat Modeling Framework. Available at: \url{https://linddun.org/} (Accessed: 18 July 2025).

Lowenstein Sandler (2025) OFAC Imposes Largest-Ever Penalty on Nonbank Financial Institution. Available at: \url{https://www.lowenstein.com/news-insights/publications/client-alerts/ofac-imposes-largest-ever-penalty-on-nonbank-financial-institution-for-egregious-and-sustained-sanctions-violations-a-216m-warning-to-us-fund-managers-know-your-investors-aml} (Accessed: 18 July 2025).

McKinsey (2025) From ripples to waves: The transformational power of tokenizing assets. Available at: \url{https://www.mckinsey.com/industries/financial-services/our-insights/from-ripples-to-waves-the-transformational-power-of-tokenizing-assets} (Accessed: 18 July 2025).

Microsoft (2025a) Microsoft Security Development Lifecycle Threat Modelling. Available at: \url{https://www.microsoft.com/en-us/securityengineering/sdl/threatmodeling} (Accessed: 18 July 2025).

Microsoft (2025b) Microsoft Threat Modeling Tool. Available at: \url{https://learn.microsoft.com/en-us/azure/security/develop/threat-modeling-tool} (Accessed: 18 July 2025).

Mitrade (2025) Infrastructure attacks caused a loss of \$2.1 billion in 2025. Available at: \url{https://www.mitrade.com/au/insights/news/live-news/article-3-919995-20250627} (Accessed: 18 July 2025).

NIST (2018) IR 8202, Blockchain Technology Overview. Available at: \url{https://csrc.nist.gov/pubs/ir/8202/final} (Accessed: 18 July 2025).

NIST SP 800-218 (2022) Secure Software Development Framework (SSDF) Version 1.1: Recommendations for Mitigating the Risk of Software Vulnerabilities. Available at: \url{https://csrc.nist.gov/pubs/sp/800/218/final} (Accessed: 18 July 2025).

Hypernative (2025) Hypernative Platform: Proactive Blockchain Security. Available at: \url{https://www.hypernative.io/} (Accessed: 18 July 2025).

OWASP (2021a) OWASP Top Ten. Available at: \url{https://owasp.org/www-project-top-ten/} (Accessed: 18 July 2025).

OWASP (2021b) OWASP Threat Dragon. Available at: \url{https://owasp.org/www-project-threat-dragon/} (Accessed: 18 July 2025).

OWASP (2022) OWASP SAMM. Available at: \url{https://owasp.org/www-project-samm/} (Accessed: 18 July 2025).

OWASP (2023) OWASP API Security Project. Available at: \url{https://owasp.org/www-project-api-security/} (Accessed: 18 July 2025).

OWASP (2025c) OWASP Smart Contract Top 10. Available at: \url{https://owasp.org/www-project-smart-contract-top-10/} (Accessed: 18 July 2025).

Reuters (2024) Losses from crypto hacks jump to \$2.2 bln in 2024. Available at: \url{https://www.reuters.com/technology/losses-crypto-hacks-jump-22-bln-2024-report-says-2024-12-19/} (Accessed: 18 July 2025).

Rubinstein, C. (2024) Top Cyber Threats To Watch Out For In 2025. Available at: \url{https://www.forbes.com/sites/carrierubinstein/2024/12/30/top-cyber-threats-to-watch-out-for-in-2025/} (Accessed: 18 July 2025).

Saad, M., Spaulding, J., Njilla, L., Kamhoua, C., Shetty, S., Nyang, D. and Mohaisen, D. (2019) Exploring the attack surface of blockchain: A systematic overview. arXiv preprint arXiv:1904.03487. Available at: \url{https://arxiv.org/abs/1904.03487} (Accessed: 18 July 2025).

Shostack, A. (2014) Threat modeling: Designing for security. John Wiley \& Sons. ISBN: 9781118809990.

Styra (2025) Open Policy Agent. Available at: \url{https://www.styra.com/open-policy-agent} (Accessed: 18 July 2025).

UcedaVelez, T. and Morana, M.M. (2015) Risk centric threat modeling: process for attack simulation and threat analysis. John Wiley \& Sons. ISBN: 978-0470500965. Available at: \url{https://www.wiley.com/en-us/Risk+Centric+Threat+Modeling%3A+Process+for+Attack+Simulation+and+Threat+Analysis-p-9780470500965} (Accessed: 18 July 2025).

U.S. Department of the Treasury (2024) National Money Laundering Risk Assessment. Available at: \url{https://home.treasury.gov/system/files/136/2024-National-Money-Laundering-Risk-Assessment.pdf} (Accessed: 18 July 2025).

Zhuang, Y., Liu, F., et al. (2020) Smart Contract Vulnerability Detection using Graph Neural Network. In Proceedings of the Twenty-Ninth International Joint Conference on Artificial Intelligence, pp.3283-3290. Available at: \url{https://www.ijcai.org/proceedings/2020/454} (Accessed: 18 July 2025).

\end{document}